\documentclass[12pt,preprint,aps,floatfix,onecolumn]{revtex4}
\usepackage{epsfig,amsmath,bbm,graphicx,color}

\long\def\comment#1{}

\begin{document}

\begin{widetext}

\noindent{\Large\bf Protein mechanical unfolding: importance of non-native interactions}

\vskip 5 mm

\noindent {\bf Maksim Kouza$^1$, Chin-Kun Hu$^{2,3}$, Hoang Zung$^4$ and Mai Suan Li$^{1,a)}$}

\vskip 2 mm

\noindent{$^1$Institute of Physics, Polish Academy of Sciences,
Al. Lotnikow 32/46, 02-668 Warsaw, Poland\\
$^2$Institute of Physics, Academia Sinica, Nankang,
Taipei 11529, Taiwan\\
$^3$Center for Nonlinear and Complex Systems and Department of
Physics, Chung Yuan Christian University, Chungli 32023, Taiwan\\
$^4$Computational Physics Lab, Vietnam National University\\
Ho Chi Minh city, 227 Nguyen Van Cu, Dist. 5, Vietnam}

\end{widetext}

\vskip 5 mm \noindent{\bf ABSTRACT

%\begin{abstract}
Mechanical unfolding of the fourth domain of
 {\em Distyostelium discoideum} filamin (DDFLN4) was studied
by all-atom
molecular dynamics simulations,
using the GROMOS96 force field 43a1 and the simple point charge explicit
water solvent. Our study reveals an important role 
of non-native interactions in the unfolding process. Namely, 
the existence of a peak centered at the end-to-end
extension $\Delta R \sim 22$ nm in the force-extension curve,
is associated with breaking of non-native hydrogen bonds.
Such a peak
has been observed 
in experiments but not in Go models, where non-native 
interactions are neglected.
We predict that
an additional peak occurs at
$\Delta R \sim 2$ nm 
using not only GROMOS96 force field 43a1 but also Amber 94 and OPLS force
fields.
This result would stimulate further experimental
studies on elastic properties of DDFLN4.}
%\end{abstract}

%\maketitle

\vspace{4cm}

$^{a)}$ Electronic email: masli@ifpan.edu.pl

\clearpage

\section{Introduction}

The last ten years have witnessed an intense activity in single molecule
force spectroscopy experiments in detecting
inter and
intramolecular forces of biological systems to understand
their functions and structures.
Much of the research has been focused
on elastic properties of proteins, DNA, and RNA, i.e, their
 response to an external force, following the seminal papers by
 Rief {\em et al.} \cite{Rief_Science97}, and Tskhovrebova {\em et al.} \cite{Tskhovrebova_Nature97}.
The main advantage of this technique is its ability
to separate out fluctuations of individual molecules from the
ensemble average behavior observed in  traditional bulk biochemical
experiments. This allows one for studying unfolding
pathways in detail using the end-to-end distance as a good reaction
coordinate. Moreover, the single molecule
force spectroscopy can be used to decipher the unfolding
free energy landscape (FEL) of biomolecules
\cite{Bustamante_ARBiochem_04,MSLi_JCP08}.

As cytoskeletal proteins, large actin-binding proteins play a key role
in cell organization, mechanics and signaling \cite{Stossel_NRMCB01}.
During the permanent cytoskeleton reorganization, all
involved participants are subject to mechanical stress. One
of them is the fourth domain of
 {\em Distyostelium discoideum} filamin (DDFLN4),
which binds different components of
actin-binding protein. Therefore, understanding the mechanical response of
DDFLN4 protein  to an external force is of great interest.
In recent Atomic Force Microscopy (AFM) experiments \cite{Schwaiger_NSMB04,Schwaiger_EMBO05},
Schwaiger {\em et al.}  have shown that DDFLN4 (Fig. 1),
%(Fig. \ref{native_ddfln4_strands_fig}),
 which has seven $\beta$-strands in the native state (NS),
unfolds via intermediates.
In the force-extension curve, they observed two peaks at the
end-to-end extension $\Delta R \approx 11$ and 22 nm ({\em Material and
Methods}).
However, using a Go model \cite{Clementi_JMB00},
 Li {\em et al.} \cite{MSLi_JCP08}
  have also obtained two peaks but they are located at
$\Delta R \approx 1.5$ and 11 nm
(see also {\em Material and Methods}).
 A natural question to ask is if the disagreement
between experiments and theory is due to over-simplification of 
the Go modeling, where non-native interactions between residues are omitted.
To answer this question, we have performed
all-atom
molecular dynamics simulations,
using the GROMOS96 force field 43a1 \cite{Gunstren_96}
and the simple point charge (SPC) explicit
water solvent \cite{Berendsen81}.
In order to check robustness of our results, limited all-simulations
have been carried out with the help of the Amber 94 \cite{Amber94_ff}
and OPLS \cite{OPLS_ff} force fields.

%%FIG. 1
\begin{figure}
\epsfxsize=5in
\vspace{0.2in}
%\centerline{\epsffile{native_ddfln4_strands_.eps}}
\centerline{\epsffile{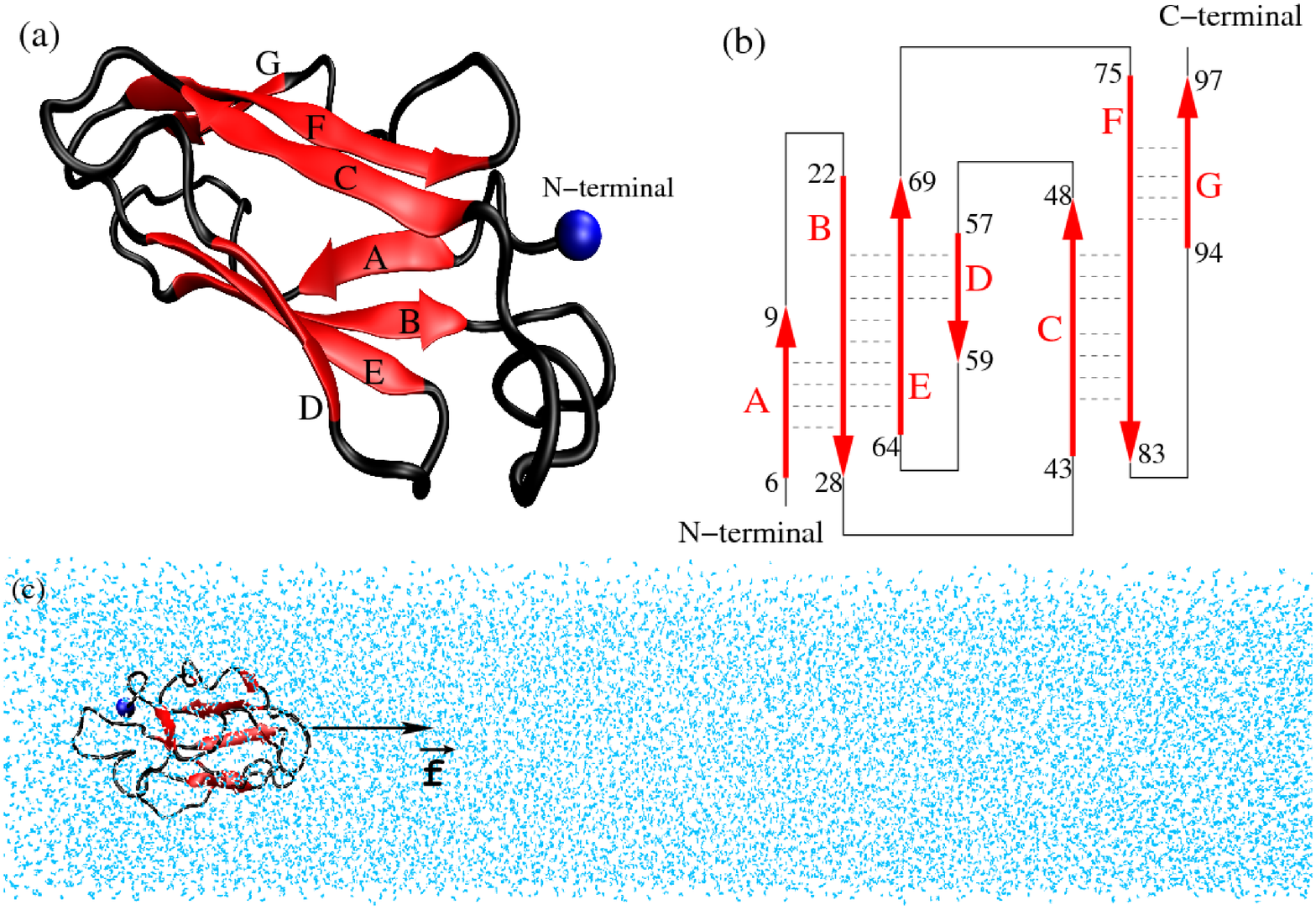}}
\caption{(a) The native conformation of DDFLN4 (PDB code: 1KSR) with
seven $\beta$-strands (A - G). (b) Schematic plot which shows
backbone contacts (dotted lines) between $\beta$-strands.
All adjacent $\beta$-strands are anti-parallel to each other.
 A, B, E, and D  belong to the first $\beta$-sheet,
whereas the second one contains  C, F, and G.
(c) The solvated system in the orthorhombic box of water (cyan).
VMD software \cite{VMD}
 was used for a plot. }
\label{native_ddfln4_strands_fig}
\end{figure}

We have shown that, 
two peaks do appear at almost
the same positions as in the experiments \cite{Schwaiger_NSMB04,Schwaiger_EMBO05}
 and more importantly, the peak at 
$\Delta R \approx 22$ nm comes from the non-native interactions.
This explains why it has not been seen in the previous Go simulations \cite{MSLi_JCP08}.
 In our opinion, this result is very
important as it opposes to the common belief \cite{West_BJ06,MSLi_BJ07a}
 that mechanical unfolding
properties of proteins are governed by their native topology.
In addition, to two peaks at large $\Delta R$, in agreement with the
Go results \cite{MSLi_JCP08},
 we have also observed a maximum
at $\Delta R \approx 2$ nm.  Because such a peak
was not detected by the AFM experiments \cite{Schwaiger_NSMB04,Schwaiger_EMBO05},
further experimental and theoretical studies are required to clarify this point.

\section{Materials and methods}

\noindent
{\bf Simulation details.} We used the GROMOS96 force field 43a1 \cite{Gunstren_96}
  to model
DDFLN4 which has 100 amino acids, and the SPC water model
\cite{Berendsen81}
 to describe the solvent. The Gromacs version 3.3.1
has been employed.
The protein was placed in an orthorhombic  box with the
edges of 4.0, 4.5 and 43 nm, and with 25000 - 26000 water molecules
(Fig. \ref{native_ddfln4_strands_fig}c).

In all simulations, the GROMACS program suite
\cite{Berendsen_CPC95,Lindahl01}
 was employed.  The equations of motion were integrated by
using a leap-frog algorithm with a time step of 2 fs.
The LINCS \cite{Hess_JCC97} was used to constrain bond lengths
with a relative geometric tolerance of $10^{-4}$.
% Covalent bond
%lengths were constrained via the  SHAKE \cite{Ryckaert77} procedure
%with a relative geometric tolerance of $10^{-4}$.
We used the
particle-mesh Ewald method to treat the long-range electrostatic
interactions \cite{Darden93}.
 The nonbonded interaction pair-list were
updated every 10 fs, using a cutoff of 1.2 nm.

The protein was minimized using the steepest decent
method. Subsequently, unconstrained molecular dynamics
simulation was performed to equilibrate the solvated system  for 100 ps
at constant pressure (1 atm) and temperature $T=300$ K with the help of
the Berendsen coupling procedure \cite{Berendsen84}.
The system was then equilibrated further at
constant temperature $T$ = 300 K and constant volume.
Afterward, the N-terminal was kept fixed and the
 force was applied to the C-terminal through a virtual cantilever moving
 at the constant velocity $v$ along the biggest $z$-axis of simulation box.
We have also performed limited simulations for the case when the N-terminal 
is pulled.

 During the simulations, the spring constant was chosen
 as $k=1000$ kJ/(mol*nm$^2) \approx 1700 $ pN/nm which is an upper
limit for $k$ of a
 cantilever used in AFM experiments.
 Movement of the pulled termini causes an extension of
 the protein and the total force can be measured by $F=k(vt-x)$,
where $x$ is the displacement of the pulled atom from its original position.
 The resulting force is computed for each time step to generate a
 force extension profile, which has peaks showing the most mechanically
 stable places in a protein.

Overall, the simulation procedure is similar to the experimental one,
except that pulling speeds in our simulations
 are several orders of magnitude higher than those used in experiments.
In the N-terminal fixed case,
we have performed simulations for
$v= 10^{9}, 5\times 10^{9}, 1.2\times 10^{10}$,
and $2.5\times 10^{10}$ nm/s, while in the AFM experiments one used
$v \sim 100 - 1000$ nm/s \cite{Schwaiger_NSMB04}.
For each value of $v$ we have generated 4 trajectories.
In the C-fixed case, the simulations were carried out for 
$v=5\times 10^{9}$ nm/s and three trajectories.

A backbone contact between amino acids $i$ and $j$ ($|i-j| > 3$)
is defined as formed if the distance between
 two corresponding C$_{\alpha}$-atoms
is smaller than a cutoff distance $d_c=6.5$ \AA .
With this choice, the molecule has 163 native contacts.
A hydrogen bond is formed provided
 the distance between donor D (or atom N) and
acceptor A (or atom O)  $\leq 3.5 \AA \,$ and the angle D-H-A
$\ge 145^{\circ}$.

The unfolding process was studied by monitoring the dependence
of numbers of backbone  contacts and hydrogen bonds (HBs)
formed by seven $\beta$-strands enumerated as
A to G (Fig. 1a)
%(Fig. \ref{native_ddfln4_strands_fig}a) 
 on the end-to-end extension.
In the NS, backbone contacts exist between seven pairs
of $\beta$-strands
P$_{\textrm{AB}}$,
P$_{\textrm{AF}}$, P$_{\textrm{BE}}$,
P$_{\textrm{CD}}$, P$_{\textrm{CF}}$, P$_{\textrm{DE}}$, and P$_{\textrm{FG}}$
as shown in Fig. 1b.
%Fig. \ref{native_ddfln4_strands_fig}b. 
Additional information on unfolding pathways was also obtained
from the evolution of contacts 
of these pairs. To understand the nature of the third peak in the 
force-extension curve, we have also monitored the evolution of
contacts formed by non-native $\beta$-strands which emerge as the unfolding
progresses.

\begin{figure}
\epsfxsize=5in
\vspace{0.2in}
%\centerline{\epsffile{f_r_gromacs_go.eps}}
\centerline{\epsffile{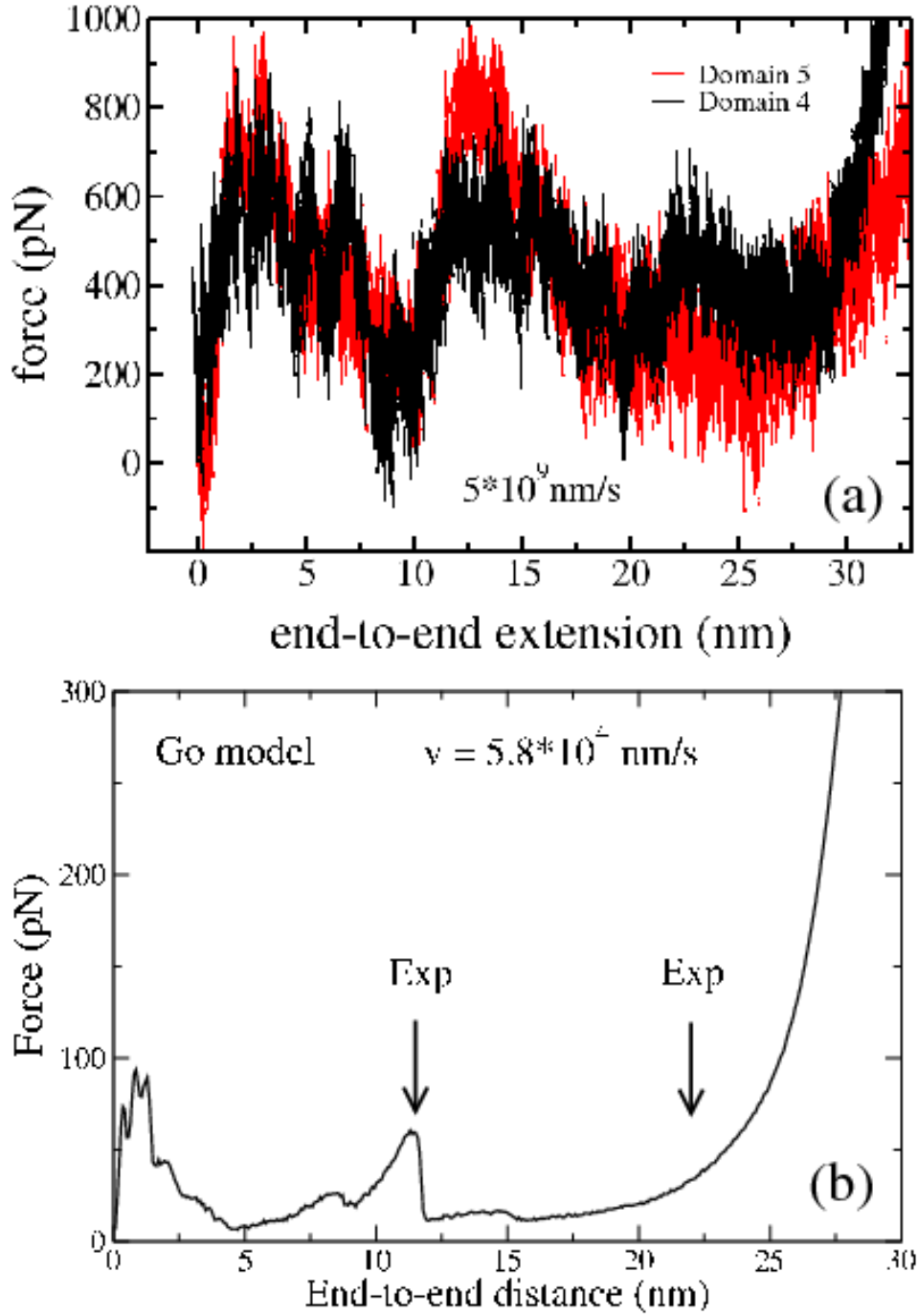}}
\caption{(a) The force-extension profiles for Domain 4 (black) and Domain 5 (red).
The results were obtained by the all-atom simulations for
the pulling speed $v=5\times 10^{9}$ nm/s.
(b) The force-extension curve for DDFLN4 and
pulling speed
$v=5.8\times 10^4$ nm/s was obtained by
 using the Go model \cite{Clementi_JMB00}.
Results were averaged over 40 trajectories.
Arrows refer to positions of two peaks
($\Delta R =11$ and 22 nm) which are expected to be the same
as observed on the experiments
\cite{Schwaiger_NSMB04,Schwaiger_EMBO05} .}
\label{f_r_gromacs_go_fig}
\end{figure}

\noindent
{\bf Validity of GROMOS96 force field 43a1}.
It should be noted that the Gromos force field has been proved
useful for studying structural dynamics and kinetics of peptides and proteins
\cite{Klepeis_COSB09}. It can not only decipher folding
mechanisms but also provide reasonable estimates for folding times \cite{Nguyen_Proteins05}.
While the steered molecular dynamics with NAMD \cite{Phillips_JCC05} is
 widely
used for studying mechanical properties of biomolecules for a decade, 
the pulling option has been recently implemented in
the GROMACS software. Therefore,
we want to check its validity for stretching proteins first.
Having used speed $v=5\times 10^{9}$ nm/s, we can demonstrate
(results not shown) that
the force-extension profile for the domain I27 is similar to the result
obtained by NAMD \cite{Lu_BJ98}. We have also carried out
limited runs for domain 5 (DDFLN5) for comparison with DDFLN4.
In agreement with the experiments \cite{Schwaiger_NSMB04,Schwaiger_EMBO05},
domain 4 is less stable and has one peak more than DDFLN5
(Fig. \ref{f_r_gromacs_go_fig}a).
Thus, the GROMACS software with the GROMOS96 force field 43a1 can serve as a reliable
tool for studying protein mechanical unfolding.
To ensure that this conclusion is correct, we have also performed
a limited set of simulations using all-atom Amber 94 \cite{Amber94_ff} and 
OPLS \cite{OPLS_ff} force fields. The simulations were carried out
within the GROMACS program suite
\cite{Berendsen_CPC95,Lindahl01}.

% FIG. 2

\noindent
{\bf A short survey on previous
results obtained by experiments and Go simulations.}
In order to make the presentation transparent, let us briefly discuss the previous experimental
\cite{Schwaiger_NSMB04,Schwaiger_EMBO05} and simulation results
\cite{MSLi_JCP09}. Experimentally one observed two peaks
in the force-extension curve \cite{Schwaiger_NSMB04,Schwaiger_EMBO05}.
However, determining their precise location as a function of the end-to-end
distance
is not an easy task, because, on experiments, unfolding lengths were
measured by fitting data to the worm-like chain model
but not by direct peak spacing.
Schwaiger {\em et al.} \cite{Schwaiger_NSMB04} reported that the
first peak occurs
due to unfolding of strands A and B and the loop between
B and C. In this first unfolding
event, the length changes by $\Delta L_1 = 14 -15$ nm,
 where $L$ is the length
parameter in the worm-like chain model.
Assuming the distance
 between to neighboring amino acides $a=0.36$ nm, they
have shown that this length gain corresponds to full stretching
of  $\sim$ 40 first
residues. Since $L$ and $R$ have different physical meanings, this does
not mean that the first peak is located at the extension
 $\Delta R = \Delta L_1$. Here, we propose to estimate its location
by comparing the experimental data with our simulation results
\cite{MSLi_JCP09}. Note that such a trick has been already used to
locate the position of the hump in the force-extension curve for the
titin domain I27 \cite{Lu_BJ98,Marszalek_Nature99}.
Using the Go model \cite{Clementi_JMB00} and the pulling speed
$v=5.8\times 10^4$ nm/s, we obtain the the force-extension curve
shown in Fig. \ref{f_r_gromacs_go_fig}b. 
The first peak at $\Delta R \approx 1.5$ nm in this theoretical curve
cannot correspond to the first peak observed in the experiments for
two reasons. First, this value of
$\Delta R$ is too small to describe unfolding of 40 first residues.
Second, the first experimental peak occurs due to unfolding of
strands A and B \cite{Schwaiger_NSMB04}, while the theoretical peak
corresponds to breaking of native contacts between  strands A and F
\cite{MSLi_JCP09}. The second theoretical peak at $\Delta R = 11 \pm$ 1 nm
(the error bar comes from averaging over different trajectories),
could be identified as the first
experimental peak because it also corresponds to unfolding of strands A and B
\cite{MSLi_JCP09}.
Thus, assuming that the Go model correctly captures the first experimental
peak, one can infer that this peak is located at $\Delta R = 11 \pm$ 1 nm.
From the experimental data \cite{Schwaiger_NSMB04,Schwaiger_EMBO05}, we can
estimate the distance between two peaks $\sim 11 \pm 1$ nm.
Therefore the second experimental peak is expected to locate at
$\Delta R= 22\pm 2$ nm.
This peak was not observed in our previous Go simulations
(Fig. \ref{f_r_gromacs_go_fig}b) \cite{MSLi_JCP09}.
But we will show that the all-atom models, where the non-native interactions
are taken into account, can reproduce it.

\section{Results}

\subsection {Existence of three peaks in force-extension profile}

\begin{figure}
\epsfxsize=5in
\vspace{0.2in}
\centerline{\epsffile{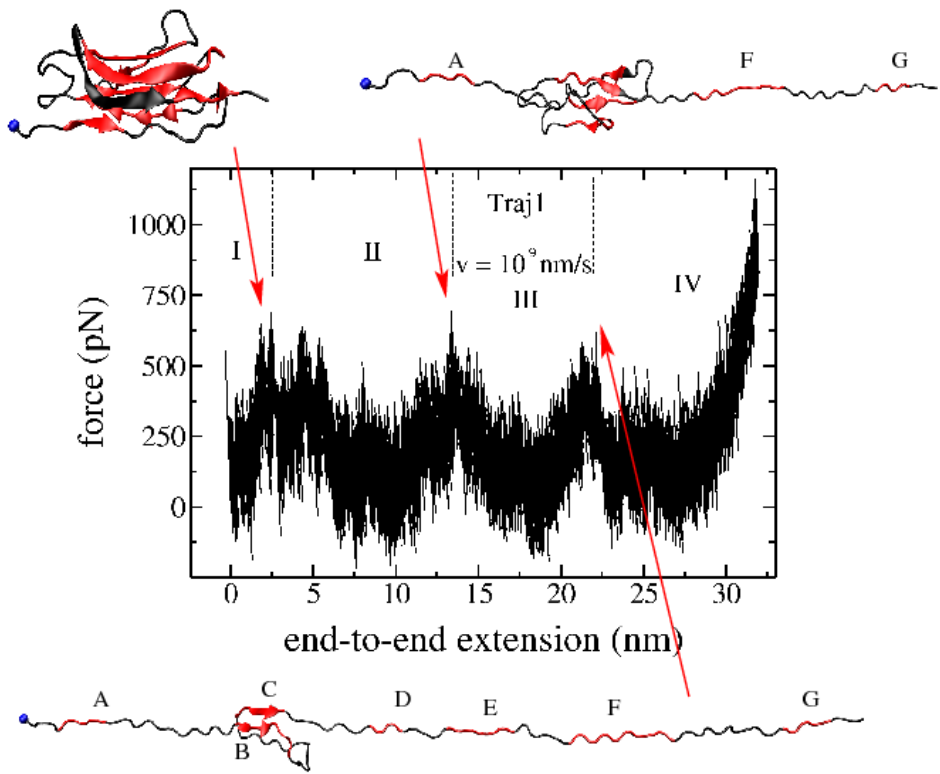}}
\caption{Force-extension profile for trajectory 1 for $v=10^9$ nm/s.
Vertical dashed lines separate four unfolding regimes.
Shown are typical snapshots around three peaks.
Heights of peaks (from left) are $f_{max1}=695$ pN, $f_{max2}=704$ pN,
and $f_{max3}=626$ pN.}
\label{fe1}
\end{figure}

\begin{figure}
\epsfxsize=6in
\vspace{0.2in}
\centerline{\epsffile{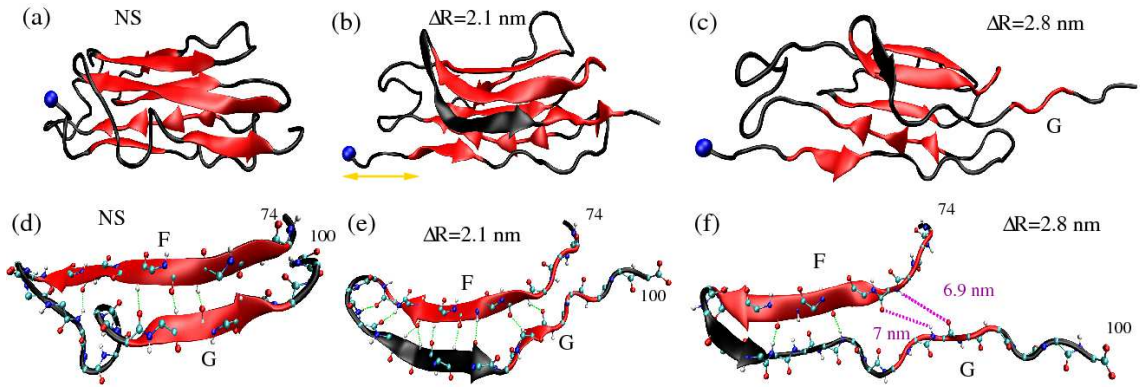}}
\caption{(a) The NS conformation is shown for comparison with
the other ones. (b) A typical conformation before the first unfolding
event takes place ($\Delta R \approx$ 2.1 nm).
The yellow arrow shows
a part of protein which starts to unfold. An additional
non-native $\beta$-strand between amino acids 87 and 92 is marked by
black color. (c)
A conformation after the first peak, at $\Delta R \approx$ 2.8 nm,
where strand G has already
detached from the core. (d) The same as in (a) but
4 HBs
(green color) between $\beta$-strands are displayed. (e) The same as in (b)
 but all 9 HBs are shown.
(f) The same as in (c) but
broken HBs (purple) between F and G  are displayed.}
\label{2nm}
\end{figure}

Since the results obtained for four pulling speeds ({\em Material and Methods})
are qualitatively similar, we will focus on the smallest 
$v=10^9$ nm/s case.
The force extension curve, obtained at
this speed, for the trajectory 1, can be
divided into four regions (Fig. \ref{fe1}):

 {\em Region I ($0 \lesssim \Delta R \lesssim 2.4$ nm)}.
Due to thermal fluctuations, the total force fluctuates a lot,
 but, in general,
it increases and reaches the first maximum $f_{max1}=695$ pN
at $\Delta R \approx$ 2.42 nm.
A typical snapshot before the first unfolding event (Fig. \ref{fe1})
 shows that structures remain native-like.
During the first period, the N-terminal part is being extended,
 but the protein maintains all $\beta$-sheet secondary structures
(Fig. \ref{2nm}b).
 Although, the unfolding starts from the N-terminal (Fig. \ref{2nm}b),
 after the first peak, strand G from
the C-termini got unfolded first (Fig. \ref{2nm}c and \ref{2nm}f).
 In order to understand the nature of this peak
on the molecular level, we consider the evolution of HBs in detail.
As a molecule departs from the NS, 
 non-native hydrogen bonds are created and at $\Delta R = 2.1$ nm, e.g.,
 a non-native $\beta$-strand between amino acids 87 and
92 (Fig. \ref{2nm}b)
 is formed. This leads to
increase of the number of HBs between F and G
from 4 (Fig. \ref{2nm}d) to 9 (Fig. \ref{2nm}e).
Structures with the enhanced number of HBs should show strong resistance to
the external perturbation and the first peak occurs due to
their unfolding (Fig. \ref{2nm}b).
It should be noted that this maximum was also observed in
the Go simulations \cite{MSLi_JCP08,MSLi_JCP09},
 but not 
in the experiments \cite{Schwaiger_NSMB04,Schwaiger_EMBO05}.
Both all-atom and Go simulations reveal that the unfolding
of strand G is responsible for its occurrence.

%%FIG .3

{\em Region II ($2.4 nm \lesssim \Delta R \lesssim 13.4$ nm):}
After the first peak, the force drops rapidly
 from 695 to 300 pN and secondary structure elements begin to break down.
During this period, strands  A, F and G unfold completely,
whereas  B, C, D and E strands remain structured 
(see Fig. \ref{fe1}
 for a typical snapshot).

%%%FIG.  4

{\em Region III ($13.4 nm \lesssim \Delta R \lesssim 22.1$ nm):}
During the second  and third stages, the complete unfolding of
strands D and E takes place. Strands
B and C undergo significant conformational changes,
losing their equilibrium hydrogen bonds. Even though a core
formed by them
remains compact (see bottom of Fig. \ref{fe1}
 for a typical snapshot).
Below we will show in detail that
the third peak is associated with breaking
of non-native hydrogen bonds between strands B and C.

%%%FIG.  5
\begin{figure}
\epsfxsize=4.5in
\vspace{0.2in}
%\centerline{\epsffile{force_ext_v2_traj23.eps}}
\centerline{\epsffile{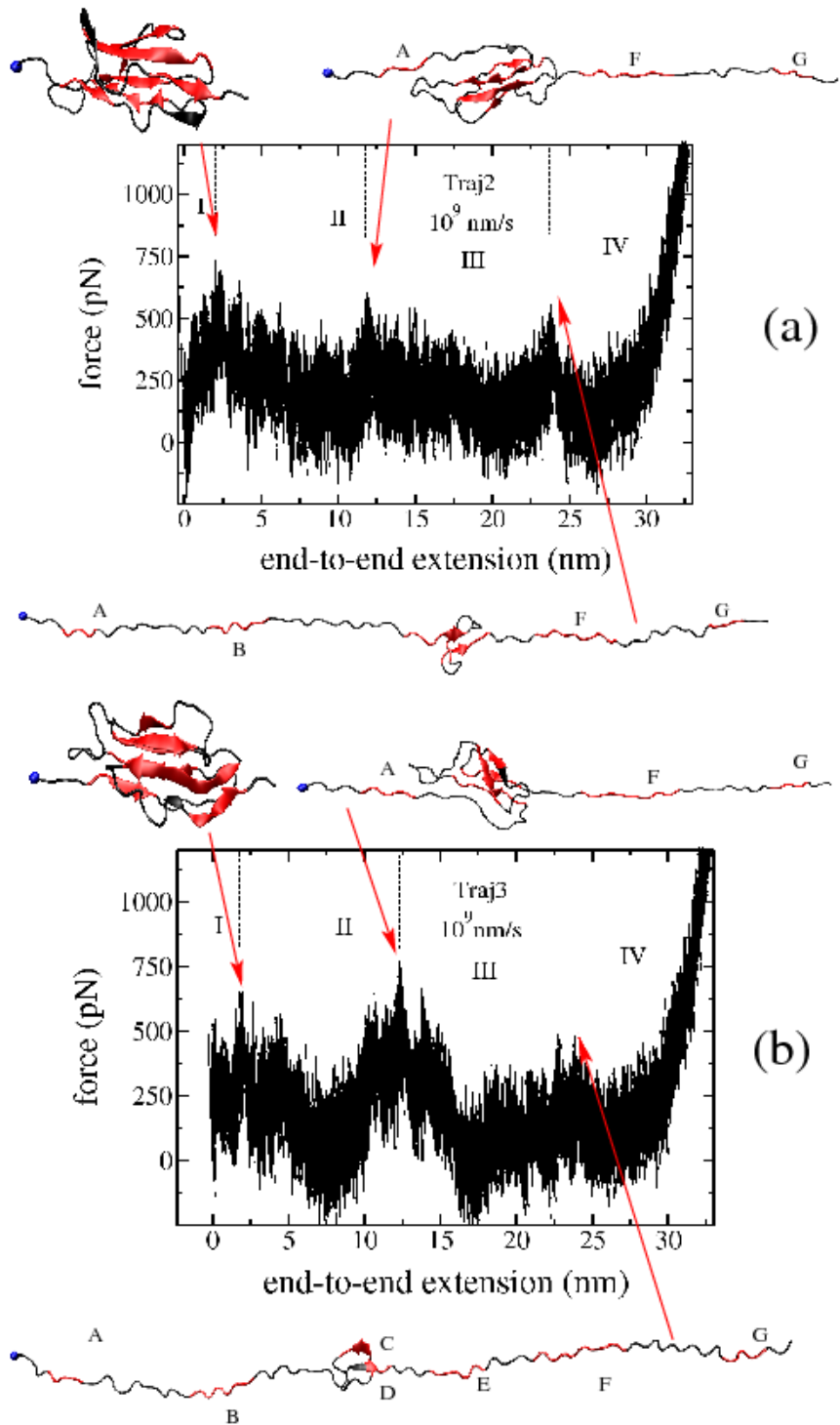}}
\caption{The same as in Fig. \ref{fe1}
but for trajectory 2 (a) and 3 (b).
Heights of peaks (from left) are $f_{max1}=740$ pN, $f_{max2}=614$ pN,
and $f_{max3}=563$ pN for trajectory 2 and $f_{max1}=685$ pN, $f_{max2}=773$ pN,
and $f_{max3}=500$ pN
 for trajectory 3.}
\label{cont_ext_traj23_fig}
\end{figure}

{\em Region IV ($\Delta R \gtrsim 22.1$ nm:)} After
breaking of non-native HBs between B and C, 
the polypeptide chain gradually reaches its rod state.

%%%FIG.  6
\begin{figure}
\epsfxsize=5in
\vspace{0.2in}
%\centerline{\epsffile{force_ext_v2_traj4av.eps}}
\centerline{\epsffile{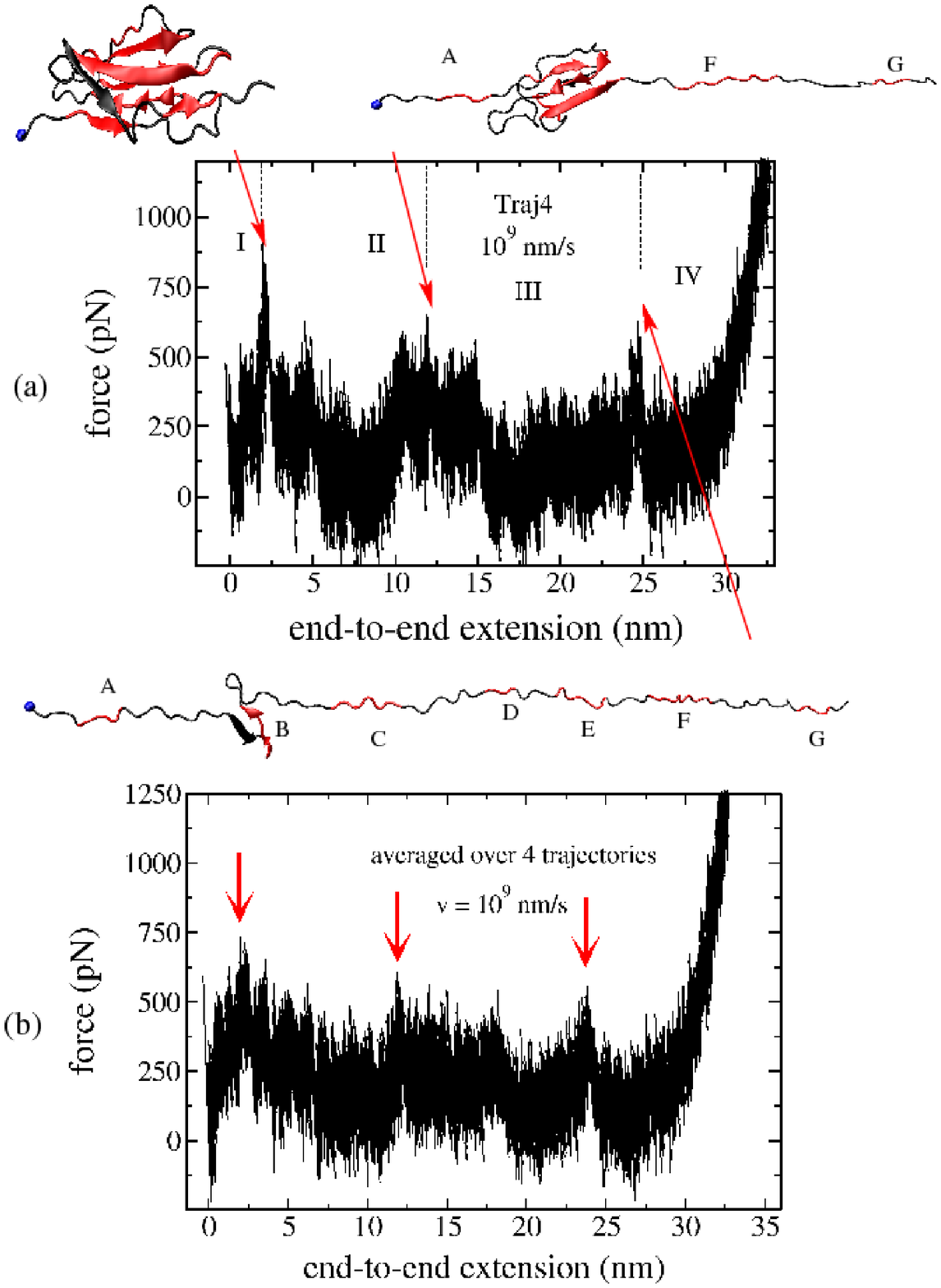}}
\caption{(a) The same as in Fig. \ref{fe1}
but for trajectory 4.
Heights of peaks (from left) are $f_{max1}=906$ pN, $f_{max2}=652$ pN,
and $f_{max3}=630$ pN. (b) The averaged over 4 trajectories
 force-extension profile at $v=10^9$ nm/s. The corresponding
 peaks are $f_{max1}=597$ pN, $f_{max2}=530$ pN,
and $f_{max3}=398$ pN. The distance between the second and third peaks is about 12 nm.}
\label{cont_ext_traj4_all_fig}
\end{figure}

For the pulling speed $v=10^9$ nm/s, the existence of three peaks
 is robust as they are also observed
in all three remaining  trajectories (Fig. \ref{cont_ext_traj23_fig}
and \ref{cont_ext_traj4_all_fig}a).
It is also clearly evident from Fig. \ref{cont_ext_traj4_all_fig}b,
 where the force-extension curve, averaged over 4
trajectories, is displayed. 
The positions of peaks fluctuate from trajectory to trajectory
but within error bars  the locations of first two peaks are in the reasonable
 agreement with the Go
simulations \cite{MSLi_JCP09}. The distance between the second
and third peaks is about 12 nm (Fig. \ref{cont_ext_traj4_all_fig}b)
and this is in accord with the experiments 
\cite{Schwaiger_NSMB04,Schwaiger_EMBO05}.

\subsection{Robustness of three peaks against pulling speeds}

\begin{figure}
\epsfxsize=4.5in
\vspace{0.2in}
%\centerline{\epsffile{fel_schem_fmax_logv_.eps}}
\centerline{\epsffile{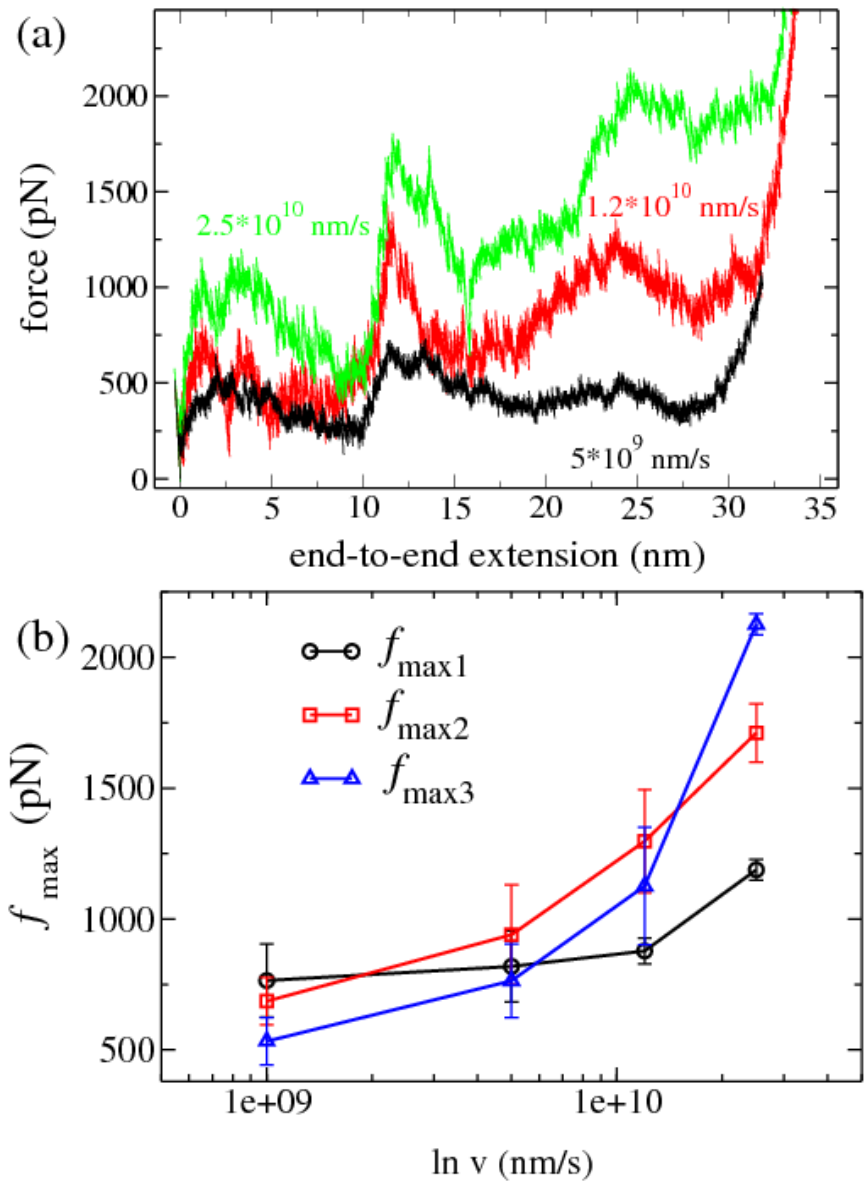}}
\caption{(a) Force-extension profiles for three values of $v$ shown next to the
curves.
(b) Dependence of heights of three peaks on $v$. Results are averaged over
four trajectories for each value of $v$.}
\label{fel_schem_fmax_logv_fig}
\end{figure}

The question we now ask is if
the existence of three peaks is independent of pulling speeds.
To this end, we performed simulations at various loading speeds
and  the results are shown in Fig. \ref{fel_schem_fmax_logv_fig}a.
In accordance with the kinetic theory \cite{Evans_BJ97}, the
heights of maxima
decrease as $v$ is lowered, but three peaks exist for all values
of pulling speeds.
Since the first peak was not observed in the experiments
\cite{Schwaiger_NSMB04,Schwaiger_EMBO05},
 a natural question emerges
is whether it is an artifact of high pulling speeds used in our simulations.
Except data at the highest value of $v$ (Fig. \ref{fel_schem_fmax_logv_fig}b),
within error bars three maxima are compatible. Therefore, in agreement
with the coarse-grained simulation results \cite{MSLi_JCP09},
the peak centered at $\Delta R \approx 2$ nm is expected to remain
at experimental loading rates.
It remains unclear if this is a shortcoming of theory or of
experiments because it is hard to imagine that a $\beta$-protein like
DDFLN4
displays
the first peak at such  a large extension $\Delta R \approx 11$ nm.
 The force-extension curve of
the titin domain I27, which has a similar native topology, for example, displays the
first peak at $\Delta R \approx 0.8$ nm
\cite{Marszalek_Nature99}.
From this prospect, our theoretical result for DDFLN4 is more favorable.
One of possible reasons of why the experiments did not detect this
maximum is related to a strong linker effect as
a single DDFLN4 domain is sandwiched between Ig domains
I27-30 and domains I31-34 from titin \cite{Schwaiger_NSMB04}.
Remember that in the case of the domain I27, the hump in
the force-extension curve was not observed in
the first experiments \cite{Rief_Science97}. It was experimentally detected  
\cite{Marszalek_Nature99} only after the theoretical prediction for its
existence \cite{Lu_BJ98}. It would be very interesting if the first peak
predicted by our simulations will be confirmed by experiments.

%%%FIG. 7

\subsection{Importance of non-native interactions}

\begin{figure}
\epsfxsize=5.5in
\vspace{0.2in}
\centerline{\epsffile{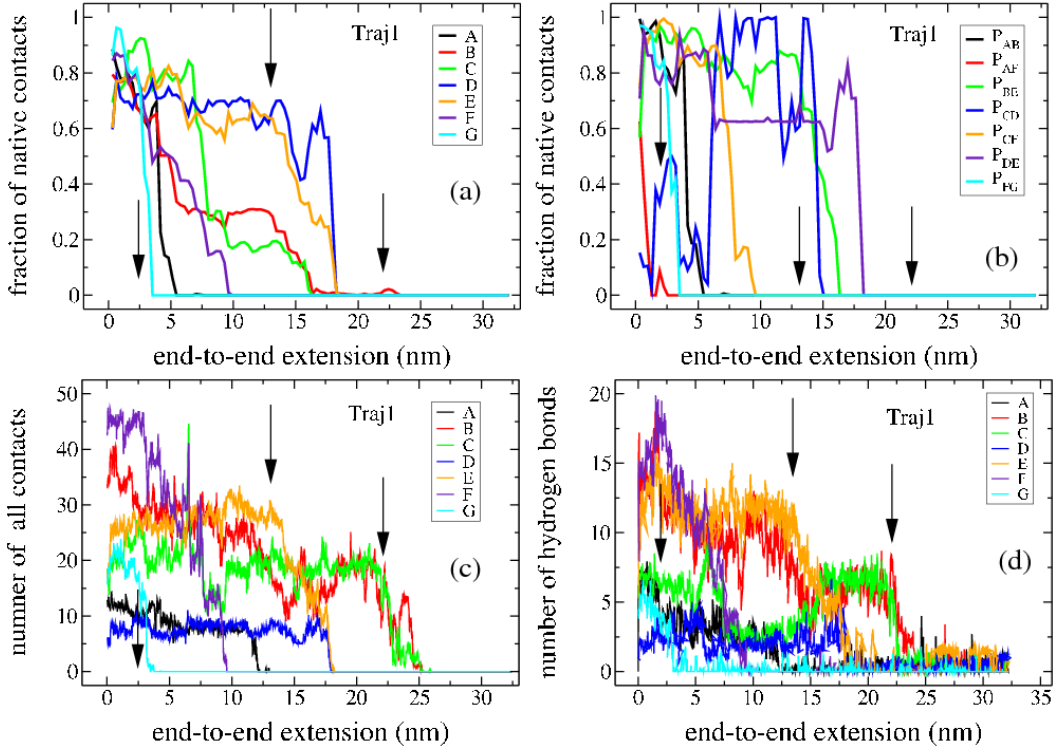}}
\caption{(a) Dependence of the number of native backbone contacts formed by
individual strands on $\Delta R$ for $v=10^9$ nm/s. Arrows refer to
positions of three peaks in the force-extension curve. (b) The same
as in (a) but for pairs of strands. (c)
The same as in (a) but for all contacts (native and non-native).
(d) The same as in (c) but for HBs.}
\label{cont_ext_traj1_fig}
\end{figure}

As mentioned above, the third peak at $\Delta R \approx 22$ nm was observed in the experiments
but not in Go models \cite{MSLi_JCP08,MSLi_JCP09},
 where 
non-native interactions are omitted.
In this section, we show, at molecular level, that these very interactions lead
to its existence. 
To this end, for the first trajectory of the lowest pulling
speed ($v=10^9$ nm/s), we plot the number of native contacts formed
by seven strands and their pairs as a function of $\Delta R$.
The first peak
corresponds to unfolding of strand G (Fig. \ref{cont_ext_traj1_fig}a)
 as all (A,F) and (F,G) contacts
are broken just after passing it (Fig. \ref{cont_ext_traj1_fig}b).
Thus, the structure of the first intermediate state,
 which corresponds to this peak, consists of 6 ordered strands A-F
(a typical snapshot is given in Fig. \ref{2nm}c).

%%%FIG.  8

The second unfolding event
is associated with full unfolding of A and F and the drastic decrease of 
native contacts of B and C (Fig. \ref{cont_ext_traj1_fig}a). 
After the second peak
only (B,E), (C,D) and (D,E) native contacts survive (\ref{cont_ext_traj1_fig}b).
The structure of the corresponding
second intermediate state  contains partially
structured strands B, C, D and E. A typical snapshot is displayed
in top of Fig. \ref{fe1}.

%%%FIG.  9

\begin{figure}
\epsfxsize=6in
\vspace{0.2in}
%\centerline{\epsffile{HB_pair_ext_traj12.eps}}
\centerline{\epsffile{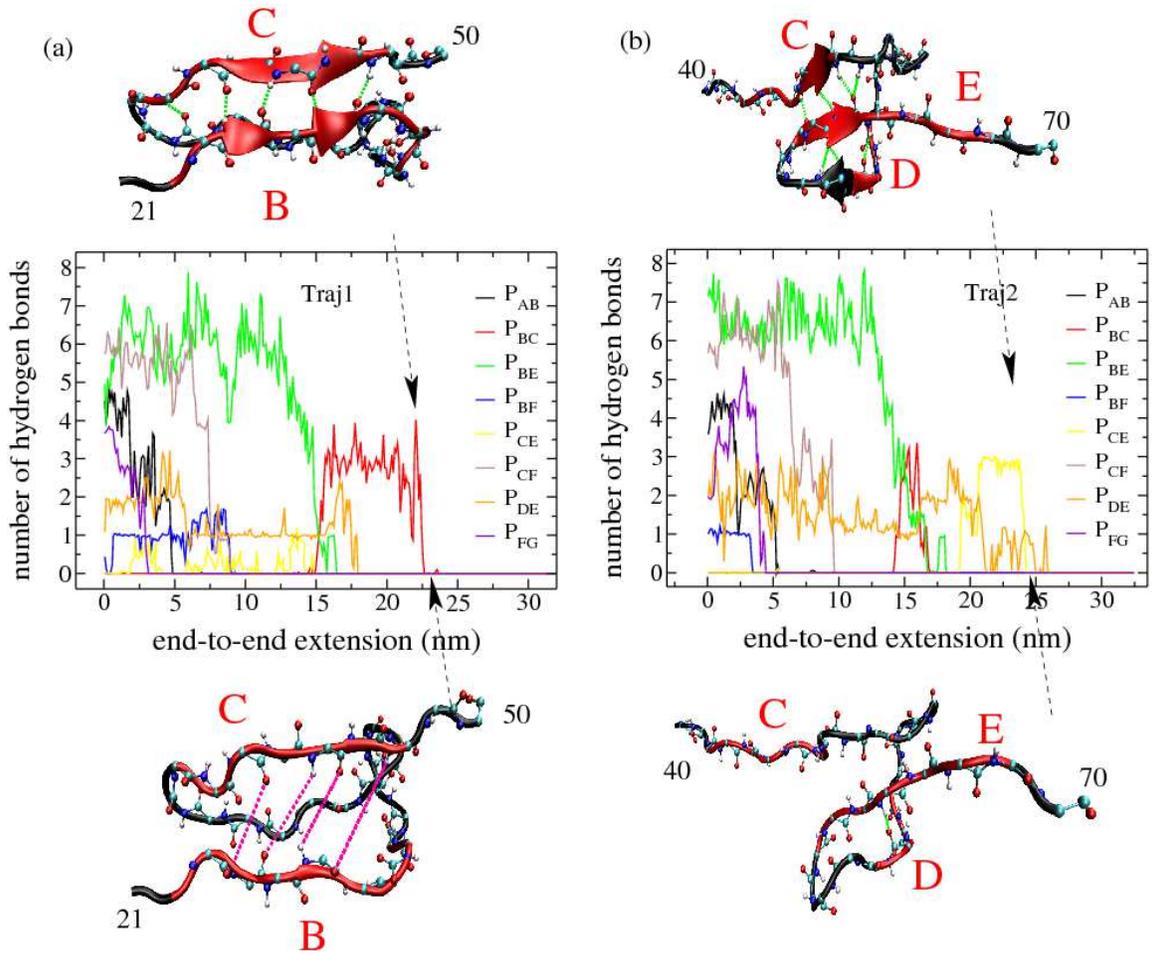}}
\caption{(a) Dependence of the number of HBs between pairs
of strands for the first trajectory and $v=10^9$ nm/s.
Non-native HBs between strands
B and C start to appear at $\Delta R \approx$ 15 nm
(red curve). Their
rupture leads to the maximum centered at $\Delta R \approx 22.4$ nm.
Upper snapshot shows five HBs between B an C
(green) before the third
unfolding event.
Lower snapshot is a fragment after the third peak,
 where all HBs are already broken
(purple dotted lines).
(b) The same as in (a),
but for trajectory 2. The upper snapshot shows the HBs
between strands
C and E before their rupture. Just after the last unfolding event,
one HB between D and E also survives (lower snapshot).}
\label{HB_pair_ext_traj12_fig}
\end{figure}

Remarkably, for $\Delta R \gtrsim 18$ nm, none of native contacts exists,
except very small fluctuations of a few contacts of strand
 B around $\Delta R \approx 22.5$ nm (Fig. \ref{cont_ext_traj1_fig}a).
Such  fluctuations are negligible as they
are not even manifested  in existence of native contacts
between corresponding pairs (A,B) and (B,E) (Fig. \ref{cont_ext_traj1_fig}b)
Therefore, we come to a very interesting conclusion that the third peak
centered at $\Delta R \approx 22.5$ nm is not related to native interactions.
This explains why it was not detected by simulations
\cite{MSLi_JCP08,MSLi_JCP09}
 using the Go model \cite{Clementi_JMB00}.

The mechanism underlying occurrence of the third peak may be revealed
using the results shown in Fig. \ref{cont_ext_traj1_fig}c, where
the number of all backbone contacts (native and non-native) 
is plotted as a function of $\Delta R$. Since,
for $\Delta R \gtrsim 18$ nm,  native contacts vanish, this peak
is associated with an abrupt decrease of non-native contacts between strands
B and C. Its nature may be also understood by monitoring
the dependence of HBs on $\Delta R$ (Fig. \ref{cont_ext_traj1_fig}d),
which shows that
the last maximum is caused by loss of
HBs of these strands.
More precisely, five HBs between B and C, which were not present in
the native conformation, are broken (Fig. \ref{HB_pair_ext_traj12_fig}a). 
Interestingly, these bonds appear at $\Delta R \gtrsim$ 15 nm, i.e.
after the second unfolding event (Fig. \ref{HB_pair_ext_traj12_fig}a).
Thus, our study can not only reproduce the experimentally observed
peak at $\Delta R \approx 22$ nm, but also shed light on its nature
on the molecular level.
From this perspective, all-atom simulations are superior to experiments.

One corollary from Fig. \ref{cont_ext_traj1_fig} is that one can not
provide a complete description of the unfolding process based on the evolution
of only native contacts. 
It is because, as a molecule extends, its
secondary structures 
 change and new non-native secondary structures may occur.
Beyond the extension of 17-18 nm (see snapshot at bottom of Fig. \ref{fe1}),
 e.g., the protein
lost all native contacts, but it does not get a 
extended state without any structures.
Therefore, a full description of mechanical unfolding may be obtained
by monitoring either  all backbone contacts or HBs,
as these two quantities give the same unfolding picture
(Fig. \ref{cont_ext_traj1_fig}c and \ref{cont_ext_traj1_fig}d).

As said above, for trajectory 1 the third peak occurs due to breaking
non-native HBs between strands B and C, but this mechanism is not
 universal for all trajectories we have studied.
In the case of second trajectory,
it  is associated with breaking of the non-native HBs between strands C and E
(Fig. \ref{HB_pair_ext_traj12_fig}b),  which have been created during the
stage III (Fig. \ref{cont_ext_traj23_fig}a). After the rupture
of the non-native HBs, a partial recovery of the  D and E strands
at $\Delta R \sim 24$ nm has been observed. However, the further unfolding of
refolded pieces of these strands does not affect the force-extension profile
significantly and its influence is much less pronounced compared non-native HBs
between C and E (Fig. \ref{HB_pair_ext_traj12_fig}b). 

%%%FIG.  10

\begin{figure}
\epsfxsize=5in
\vspace{0.2in}
%\centerline{\epsffile{HB_pair_ext_traj34.eps}}
\centerline{\epsffile{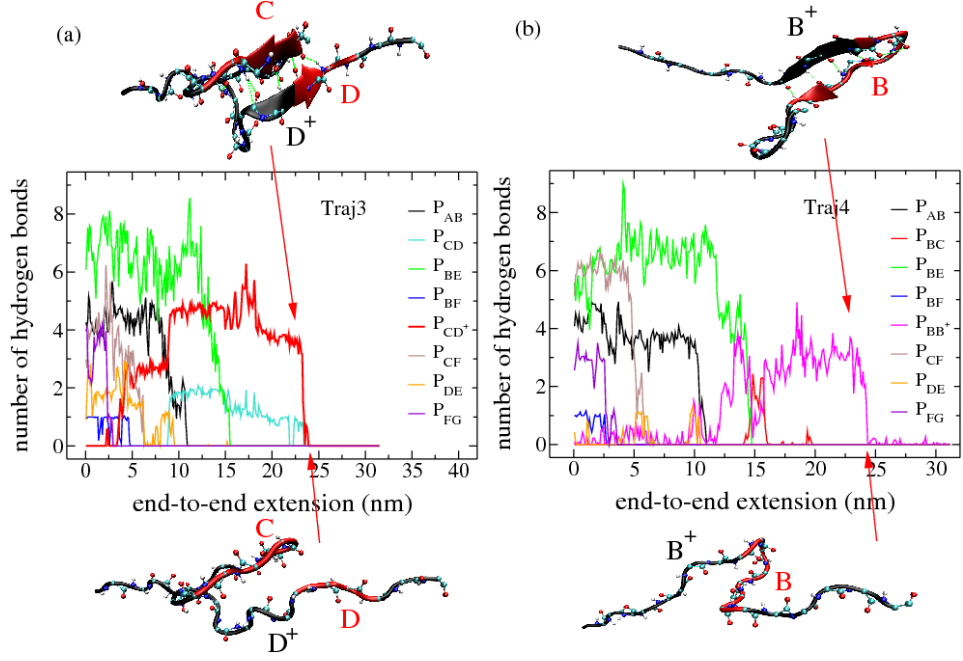}}
\caption{(a) The same as in Fig. \ref{HB_pair_ext_traj12_fig},
but for trajectory 3. The upper snapshot shows HBs between C and the non-native
strand D$^+$ (black) which includes amino acids 54-59. The rupture of these bonds (lower snapshot) leads to the occurrence
of the third peak.
(b) The same as in (a),
but for trajectory 4. Here the nature of the last peak is related to
breaking of HBs between B and the non-native strand B$^+$ (black,
amino acids 17-21)
as shown in the upper snapshot.}
\label{HB_pair_ext_traj34_fig}
\end{figure}

For the first two trajectories, it was enough to consider the evolution of
native and non-native HBs between the strands
formed in the native state. However, for the third and fourth trajectories,
determining the nature of the third peak requires consideration of
additional $\beta$-strands that appear during the unfolding process.
In the  case of trajectory 3, the native strand D has been extended for 3
 amino acids more (from amino acid 54 to 59) and this extension
is labeled as $D^+$.  In the upper snapshot of
 Fig. \ref{HB_pair_ext_traj34_fig}a, this
 non-native part is shown in black. The red curve in this figure
clearly shows that
the third peak appears due to the breaking of four non-native HBs between the
 C and $D^+$ strands. After this unfolding event HBs between all pairs are ruptured.

For trajectory 4,  HBs between all native strands were
broken before the protein reaches $\Delta R \approx 15$ nm
(Fig. \ref{HB_pair_ext_traj34_fig}b).
The nature of the third peak is purely
related to the rupture
of non-native HBs between strand B and the non-native strand $B^+$
 (from amino acid 17 to 21) which is described by the black arrow
(upper snapshot of Fig. \ref{HB_pair_ext_traj34_fig}b).
Thus , the non-native interactions are responsible for occurrence of the
last peak, but individual pathways show a rich diversity.

\subsection{Unfolding pathways: N-fixed case}

\begin{figure}
\epsfxsize=5.5in
\vspace{0.2in}
%\centerline{\epsffile{HB_pair_ext_traj2_4.eps}}
\centerline{\epsffile{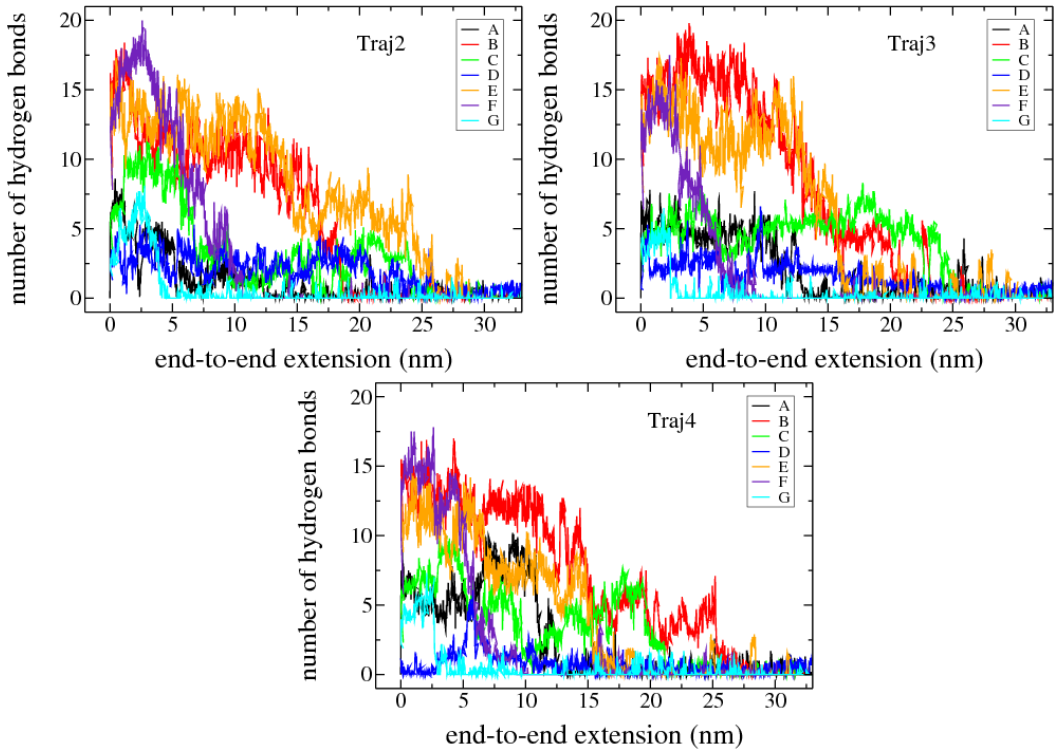}}
\caption{The end-to-end extension dependence of the
number of HBs, formed by
seven  strands, for trajectory 2, 3 and 4.
$v=10^9$ nm/s.}
\label{HB_pair_ext_traj2_4_fig}
\end{figure}

To obtain sequencing of unfolding events at $v = 10^9$ nm/s,
we use dependencies of the number of HBs on $\Delta R$.
From Fig. \ref{cont_ext_traj1_fig}d
  and  Fig. \ref{HB_pair_ext_traj2_4_fig},
 we have the following unfolding pathways for four trajectories:
\begin{eqnarray}
G \rightarrow F \rightarrow A  \rightarrow (D,E) \rightarrow (B,C), \; \;
\textrm{Trajectory 1},  \nonumber\\
G \rightarrow F \rightarrow A  \rightarrow B \rightarrow C \rightarrow (D,E),
\; \; \textrm{Trajectory 2}, \nonumber  \\
G \rightarrow F \rightarrow A  \rightarrow E \rightarrow B \rightarrow D \rightarrow
C, \; \; \textrm{Trajectory 3} \nonumber\\
G \rightarrow F \rightarrow A  \rightarrow (D,E) \rightarrow C \rightarrow B,
\; \; \textrm{Trajectory 4}.
\label{pathways_eq}
\end{eqnarray}
Although these pathways are different,
they share a common feature that the C-terminal unfolds first.
This is consistent with the results obtained by Go simulations
at high pulling speeds $v \sim 10^6$ nm/s \cite{MSLi_JCP09},
 but
contradicts to the experiments
\cite{Schwaiger_NSMB04,Schwaiger_EMBO05},
which showed that strands A and B from the N-termini unfold first.
On the other hand, our Go simulations
\cite{MSLi_JCP09}
have revealed that the agreement with the experimental results
is achieved if one performs simulations at relatively low
pulling speeds $v \sim 10^4$ nm/s.
Therefore, one can expect that the difference in
sequencing of unfolding events between present
all-atom results and the experimental ones is merely due to
large values of $v$ we used. In order to check this,
one has to carry out all-atom simulations, at least,
at $v \sim 10^4$ nm/s, but such a task is
far beyond nowaday computational facilities.

% FIGURE 11

Schwaiger {\em et al.} \cite{Schwaiger_NSMB04,Schwaiger_EMBO05} suggested that
conformations of the single unfolding intermediate which
corresponds to the first peak in the force-extension curve contain 
five strands C-G with A and B fully unstructured.
As follows from our all-atom simulations,
there exist two intermediates related to peaks located
at $\Delta R \sim$ 2 and 12 nm. Since at $\Delta R \sim$ 2 nm,
only G unfolds (Fig. \ref{2nm}c, \ref{cont_ext_traj1_fig}d
  and  \ref{HB_pair_ext_traj2_4_fig}),
the structure of the first
intermediate
consists of six structured strands A-F which is more ordered than the
experimental one.
The second intermediate, located
at $\Delta R \sim$ 12 nm, is less ordered having four strands B-E structured
(Fig. \ref{fe1}, \ref{cont_ext_traj23_fig},
\ref{cont_ext_traj4_all_fig}, \ref{cont_ext_traj1_fig}d and
\ref{HB_pair_ext_traj2_4_fig}). Again, a huge difference in pulling speeds
is a reason for differences in the nature of
experimentally and theoretically predicted intermediates.

It is very important to note that unfolding pathways depend on pulling speeds
but the number of peaks in the force-extension curve remains unchanged.
If non-native interactions are neglected as in the Go model then one has two 
peaks \cite{MSLi_JCP09}, but the inclusion of these interactions leads to
occurrence of the third maximum at $\Delta R \sim 22$ nm
(Fig. \ref{fel_schem_fmax_logv_fig}). Thus, the main meaning of the present
work is not in describing unfolding pathways but in capturing this peak.  

\subsection{Unfolding pathways: C-fixed case}

\begin{figure}
\epsfxsize=5in
\vspace{0.2in}
%\centerline{\epsffile{force_ext_Cfix_v2_total.eps}}
\centerline{\epsffile{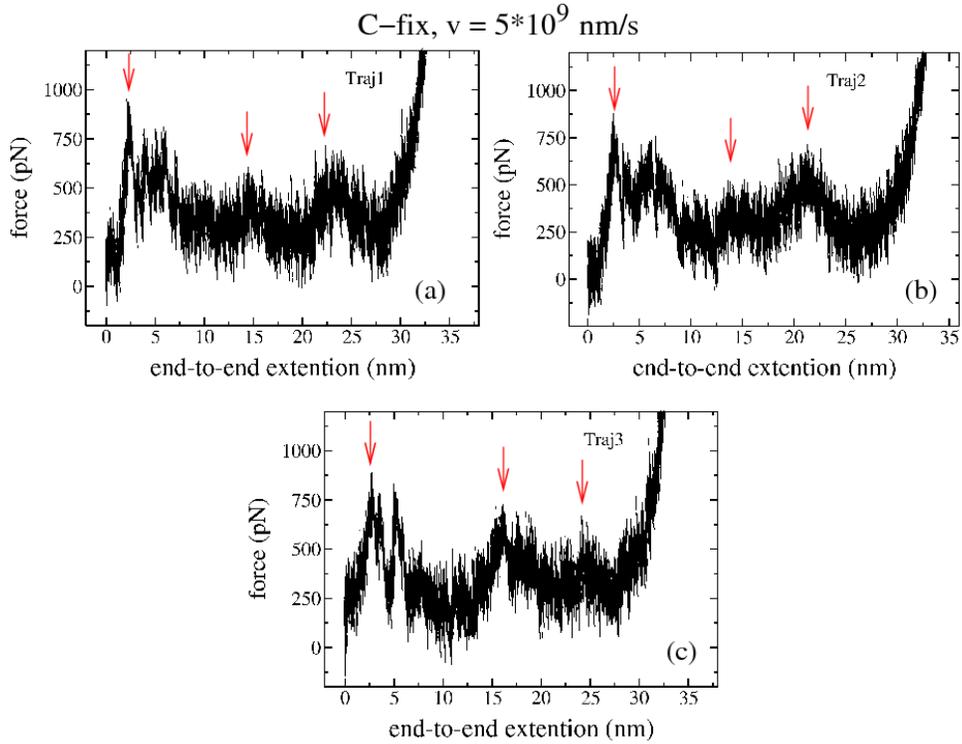}}
\caption{ The force-extension profiles, obtained in
three molecular dynamics runs with the C-terminal kept fixed.
$v=5\times 10^9$ nm/s. Arrows refer to positions of maxima.}
\label{force_ext_Cfix_v2_total_fig}
\end{figure}

It is known that unfolding pathways may depend on what end of a protein 
is kept fixed.  For ubiquitin, pulling at the
C-terminal gives pathways different from those obtained in the case
when the N-end is pulled \cite{MSLi_BJ07}. However, unfolding pathways
of the domain I27 do not depend on what terminal is fixed \cite{MSLi_BJ07}.
To study the effect of terminal fixation on unfolding pathways
of DDFLN4,
we have performed limited simulations at $v = 5\times 10^9$ nm/s
for three trajectories.
As in the N-fixed case, the force-extension curves display three peaks
(Fig. \ref{force_ext_Cfix_v2_total_fig}). Thus, the number of peaks
does not depend on what terminal is immobilized.   

%% FIG. 12

\begin{figure}
\epsfxsize=4in
\vspace{0.2in}
%\centerline{\epsffile{traj1_snapshot_Cfix.eps}}
\centerline{\epsffile{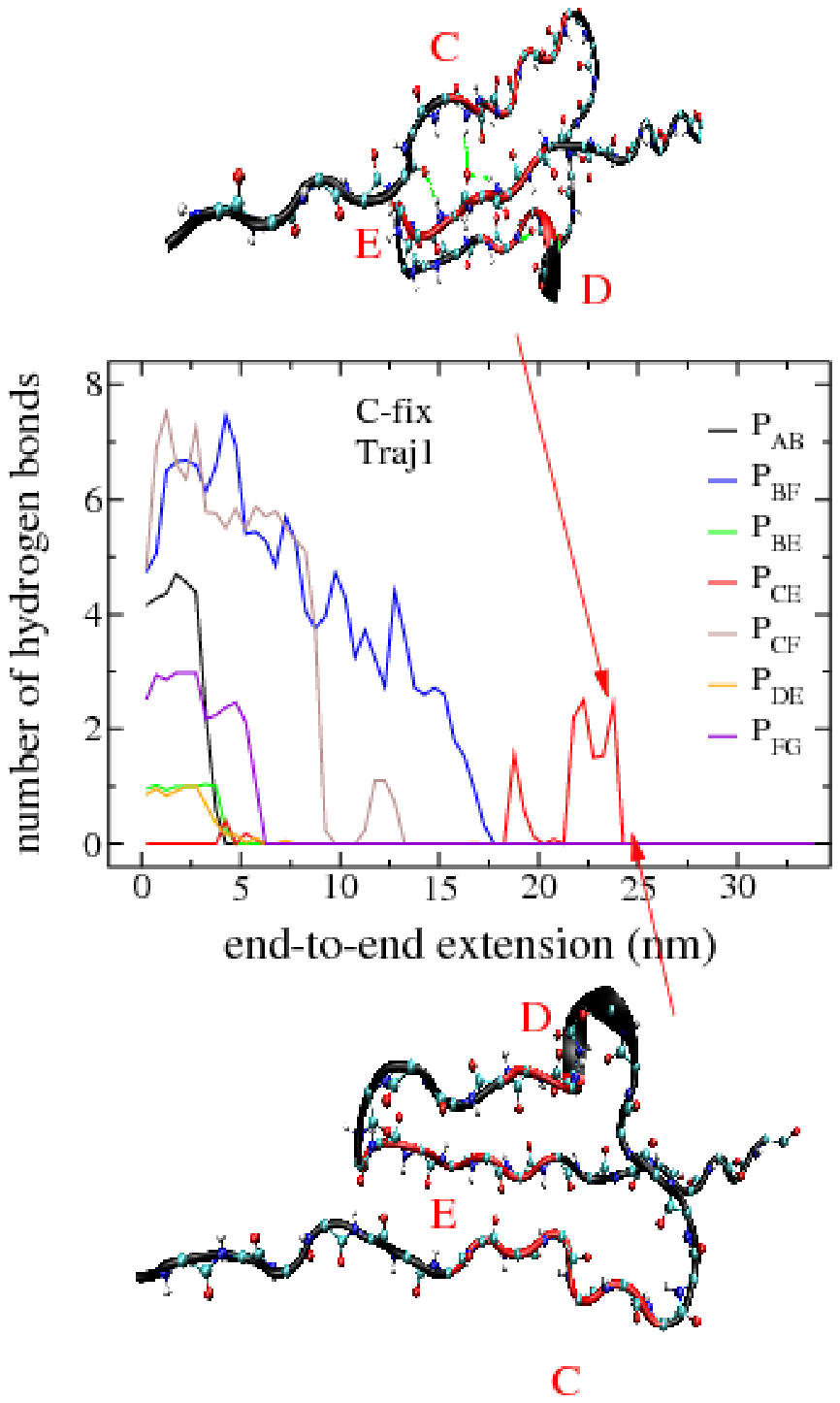}}
\caption{Dependence of the number of HBs between
pairs of strands for the first trajectory. The pulling speed
$v=5\times 10^9$ nm/s and the C-end is hold fixed. The upper snapshot
shows non-native HBs between strand C and E before their rupture.
After passing the third peak they are broken (lower snapshot).}
\label{traj1_snapshot_Cfix_fig}
\end{figure}

Fig. \ref{traj1_snapshot_Cfix_fig} shows the dependence of HBs on $\Delta R$
for trajectory 1. Before the last unfolding event, there exist only few
non-native HBs between strands C and E (upper snapshot). The lower snapshot
shows that after the third peak these bonds got ruptured. One can show that
the third peak in the second and third trajectories also comes
from the rupture of non-native HBs (results not shown). Thus, the nature
of the third peak remains essentially the same as in
the N-fixed case.

%% FIG. 13

\begin{figure}
\epsfxsize=5.5in
\vspace{0.2in}
%\centerline{\epsffile{all_contacts_Cfix_v2_total.eps}}
\centerline{\epsffile{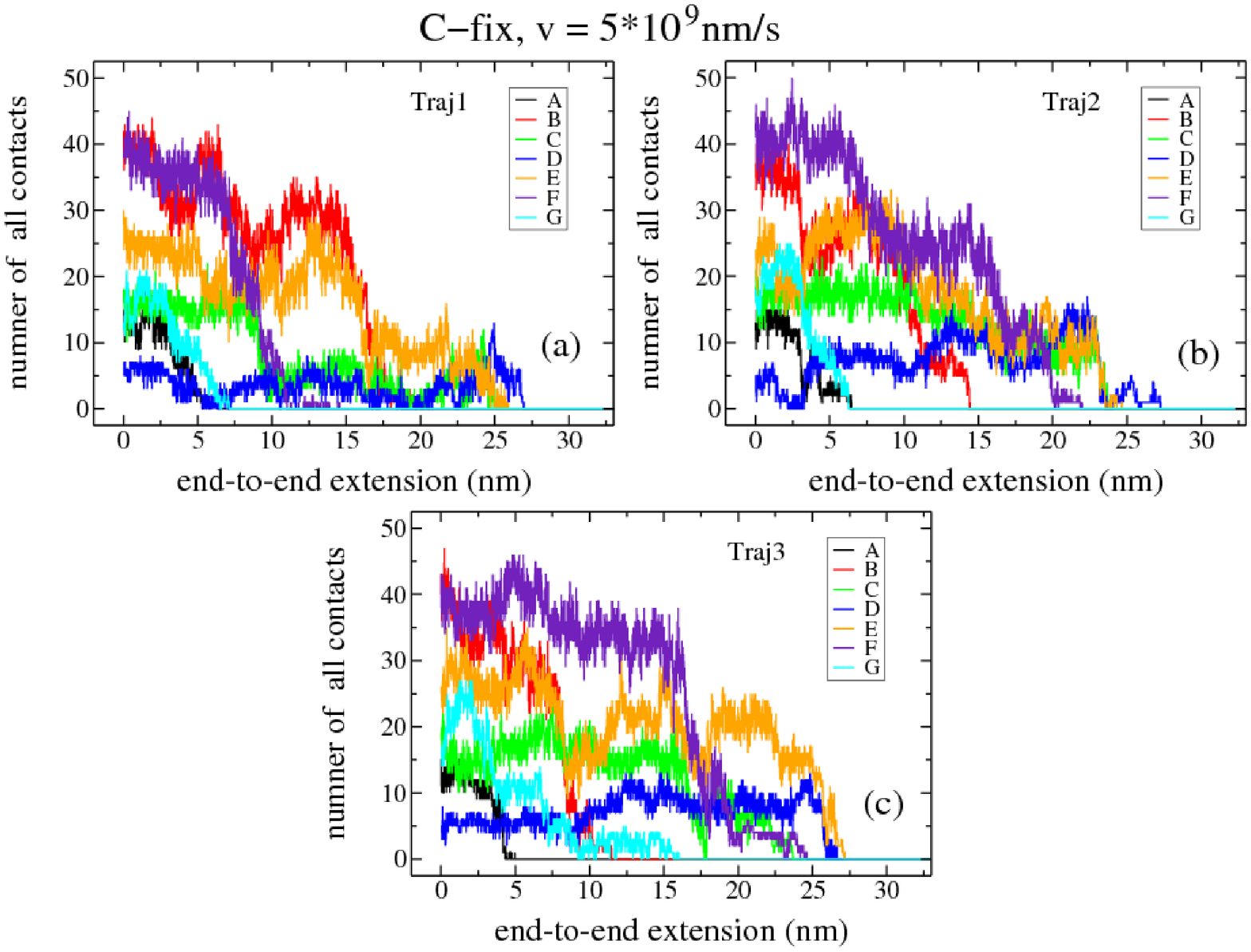}}
\caption{Dependence of the total number of all
backbone contacts (native and non-native), formed by seven strands, on $\Delta R$.
$v=5\times 10^9$ nm/s and the C-end is immobile.}
\label{all_contacts_Cfix_v2_total_fig}
\end{figure}

From the dependencies of the total number of backbone contacts formed
by secondary structures (Fig. \ref{all_contacts_Cfix_v2_total_fig}),
we obtain the following unfolding pathways:
\begin{eqnarray}
A \rightarrow G \rightarrow F  \rightarrow B \rightarrow (C,E) \rightarrow D,
\; \; \textrm{Trajectory 1}, \nonumber  \\
A \rightarrow G \rightarrow B  \rightarrow F \rightarrow (C,D,E), \; \;
\textrm{Trajectory 2},  \nonumber\\
A \rightarrow B \rightarrow G  \rightarrow (C,F) \rightarrow (E,D),
\; \; \textrm{Trajectory 3}.
\label{pathways_Cfix_eq}
\end{eqnarray}
In three cases, the strand A is unfolds first
(Eq. (\ref{pathways_Cfix_eq})). For the third trajectory
A and B unfold first and this scenario is in accord with the experiments
of Schwaiger {\em et al.} \cite{Schwaiger_NSMB04,Schwaiger_EMBO05}. However,
such an agreement is fortuitous as it happens just in one case.
As evident from Eqs . (\ref{pathways_eq}) and (\ref{pathways_Cfix_eq}),
unfolding pathways depend on what terminal is pulled. Intuitively, it is
clear because at high pulling speeds a terminal, which the external
 force is applied to,
should unfold 
first as it unfolds before the force propagates to
the opposite end.

%% FIG. 14

One can infer that the first intermediate, which corresponds to
the first peak, consists of conformations with only strand A
unfolded ( Figs. \ref{force_ext_Cfix_v2_total_fig} and
 \ref{all_contacts_Cfix_v2_total_fig}). This is valid
for all three trajectories.
Structures of the second intermediate, related to the second
unfolding event, show more diversity.
 For the first trajectory, it 
consists of 4 structured strands B-E (Fig. \ref{all_contacts_Cfix_v2_total_fig}a). After the second peak, A, F and G become unstructured.
The second and third trajectories have the same second intermediate
which contains four strands C-F (Fig. \ref{all_contacts_Cfix_v2_total_fig}b and \ref{all_contacts_Cfix_v2_total_fig}c). Overall, as in
the N-fixed case, due to high pulling speeds, the simulations give different unfolding pathways compared
to the experimental ones. Nevertheless, the third peak, which is
absent in the Go modeling, occurs in both N-fixed and C-fixed
all-atom simulations.

The question we now ask is whether the dependence of unfolding pathways
on the terminal fixation is intrinsic for DDFLN4 or this is an artifact of
high loading rates. Since it is impossible to carry out all-atom simulations
at small $v$, we have performed the simulations at $v \sim 10^4$ nm/s
using the Go model \cite{Clementi_JMB00} with the $C$-end hold fixed.
The results, obtained for the $N$-fixed case, were reported previously
\cite{MSLi_JCP09}.  It turns out that at low pulling speeds 
unfolding pathways are the same as in the the $N$-fixed case
(results not shown). Therefore, we speculate that, similar
to the domain I27 \cite{MSLi_BJ07}, at low loading rates
unfolding pathways
of DDFLN4 do not depend on the way one pulls it.
Our speculation is not unreasonable because for small $v$,
it does not matter
what end is pulled as the external force
is uniformly felt along a chain . Then, a strand, which has
a weaker
link with the core, would unfold first.

\subsection{Unfolding intermediates and pathways at low pulling speeds}

Because all-atom simulations at low $v$ are prohibited, in this section, we
try to infer information on unfolding intermediates and pathways using
the energetic argument and results followed from the Go model \cite{MSLi_JCP09}.
As shown in our previous work \cite{MSLi_JCP09},
the peak, centered at $\Delta R \approx 2$ nm, results from
breaking of contacts between A and F. Moreover, within the Go modeling,
contacts between these strands break first not only at low but also
at high loading rates. Our high-$v$ all-atom results
(Fig. \ref{cont_ext_traj1_fig}b and similar
results for other cases are not shown) also confirm this. In order to see
if (A,F) native contacts break first at low loading rates, we calculated the interaction energies
between terminal pairs (A,B), (A,F), and (F,G) using the Gromos 
force field  43a1 and equilibrium conformations.  The latter
are those conformations which have been obtained after
the equilibration step and used as initial conformations for pulling.
At the room temperature, they are
very close to the native conformation (RMSD $\approx 0.7 \, \AA$).
Since the interaction energies $E_{AB} \approx$ -120.4, $E_{AF} \approx$ -12.2,
and $E_{FG} \approx$ -114.4 kcal/mol, one can expect that (A,F)
contacts always break first. Therefore, in the first intermediate state,
all strands remain
structured but the contacts between A and F are lost.
As mentioned above, the absence of the first peak in experiments may be due to 
the strong linker effect. Another possible reason is that one needs
a relatively small amount of energy to break (A,F) couplings. This subtle
effect may be overlooked in the experiments.

At low $v$, the Go simulations \cite{MSLi_JCP09} show that A and B unfold first
and this pathway agrees very well with the experiments
of Schwaiger {\em et al.} \cite{Schwaiger_NSMB04,Schwaiger_EMBO05}.
Consequently, the corresponding theoretical intermediate, which consists of
5 strands C-G, is also identical to the experimental one. 
Since the second peak at $\Delta R \approx$ 11 nm is caused by native
interactions, its nature would be the same as in the Go 
model. Thus, we expect that at low pulling speeds, the second
intermediate provided by the Gromos force field coincides with
the experimental one.

\subsection{Robustness of the results against other force fields}

\begin{figure}
\epsfxsize=4in
\vspace{0.2in}
%\centerline{\epsffile{amber_opls_total.eps}}
\centerline{\epsffile{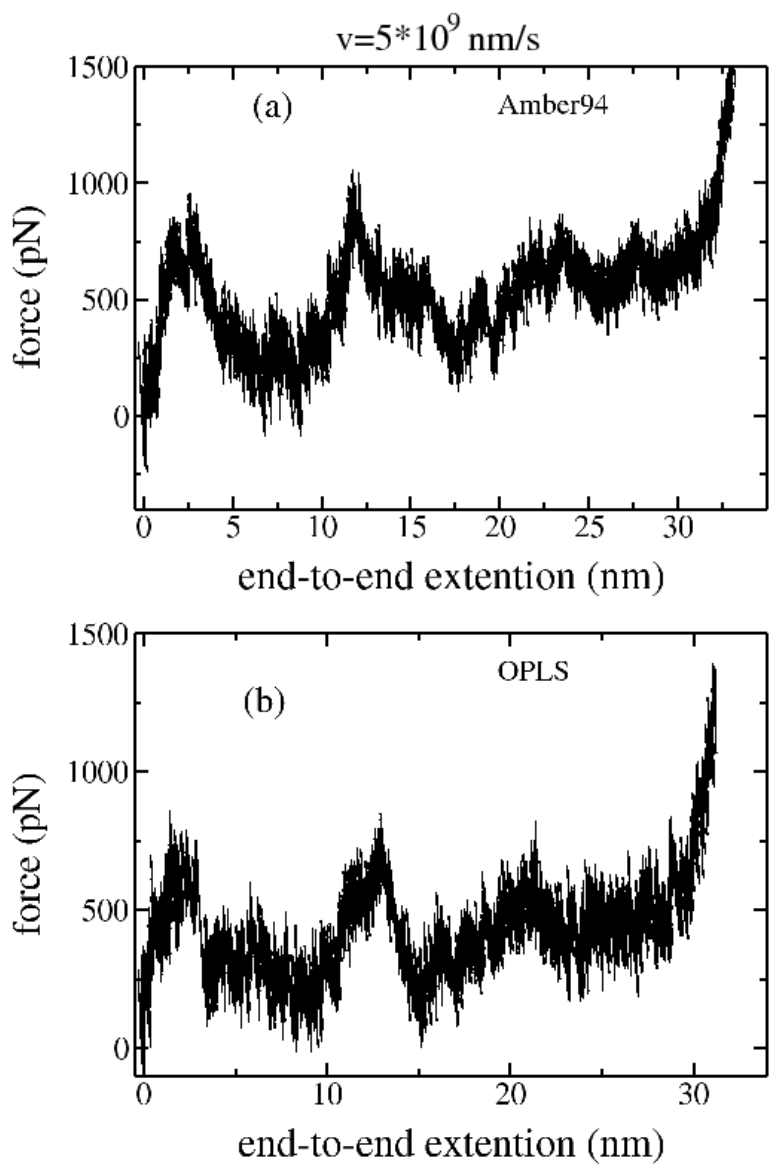}}
\caption{Shown is the force-extension profile obtained
 by using the
Amber 94 (upper panel) and OPLS (lower panel) force fields. The N-end is kept fixed and
the pulling speed $v=5\times 10^9$ nm/s.}
\label{amber_opls_total_fig}
\end{figure}

To make sure that the results obtained by the GROMOS96 force field 43a1
are robust, we have also performed a limited set of simulations for 
$v=5\times 10^9$ nm/s using the Amber 94 \cite{Amber94_ff} and 
OPLS \cite{OPLS_ff} force fields.
Fig. \ref{amber_opls_total_fig} shows two typical force-extension curves
obtained by these force fields (other
similar curves are not shown) for the pulling speed $v=5\times 10^9$ nm/s and
the N-terminus is hold fixed.
In accordance with the GROMOS96 force field 43a1 case, the Amber 94
and OPLS force fields
provide three peaks and the nature
of the third peak is also associated with non-native interactions (results not
shown). Thus, the existsence of three peaks and their nature do not depend
on the force fields we used and one can expect that this conclusion remains
valid for other existing force fields.

% FIGURE 15

\vspace {10mm}
\noindent
{\bf CONCLUSIONS}\\

We used the GROMOS96 force field 43a1 with explicit
water to understand the mechanical unfolding of the protein domain DDFLN4.
The validity of this approach was carefully checked not only by considering the
well-studied titin domain I27 and the domain DDFLN5 which is mechanically
more stable than DDFLN4,
 but also by comparison with results obtained by
Amber 94 and OPLS force fields.
We have reproduced the experimental result
on existence of two peaks located at $\Delta R \approx 11$ and 22 nm.
Our key result is that the later maximum occurs
due to breaking of non-native HBs.
It can not be encountered by the Go models in which non-native
interactions are neglected
\cite{MSLi_JCP08,MSLi_JCP09}.
Thus, our result points to the importance of these
interactions for the mechanical unfolding of DDFLN4.
For the first time we have shown that
the description of elastic properties of proteins may be not
complete ignoring non-native interactions.
This finding is valuable as the unfolding by an external
force is widely believed
to be solely governed by native topology of proteins.

Our all-atom simulation study supports the result obtained by the Go model
\cite{MSLi_JCP08,MSLi_JCP09}
 that an additional peak occurs at
$\Delta R \approx 2$ nm.
However, it was not observed by the AFM experiments
of Schwaiger {\em et al} \cite{Schwaiger_NSMB04,Schwaiger_EMBO05}.
In order to solve this controversy, one has to carry out not only
 additional experiments, but also
all-atom simulations at lower pulling speeds which are beyond
nowaday computational facilities.\\

\noindent{\bf ACKNOWLEDGMENTS}\\

The work was 
supported by 
the Ministry of Science and Informatics in Poland
(grant No 202-204-234), 
National Science Council in Taiwan
under Grant No. NSC 96-2911-M 001-003-MY3, National Center for
Theoretical Sciences in Taiwan, and Academia Sinica (Taiwan)
under Grant No. AS-95-TP-A07. MK is very grateful to the Polish committee for UNESCO
for the financial support.

\clearpage

%\bibliographystyle{biophysj}
%\bibliographystyle{proteins}
%%\bibliography{trim}
%\bibliography{References}

\end{document}